\documentclass[manuscript,screen]{acmart}
\usepackage[utf8]{inputenc}
\AtBeginDocument{%
  }

\setcopyright{acmlicensed}
\copyrightyear{2026}
\acmYear{2026}
\acmDOI{XXXXXXX.XXXXXXX}
\acmConference[Conference acronym 'XX]{Make sure to enter the correct
  conference title from your rights confirmation email}{April 13--17,
  2026}{Barcelona, Spain}
\acmISBN{978-1-4503-XXXX-X/2018/06}

\usepackage{tabularx}
\usepackage{multirow}
\begin{document}

\title{SVG Decomposition for Enhancing Large Multimodal Models Visualization Comprehension: A Study with Floor Plans}

% SVG Decomposition for Enhancing Large Multimodal Models Visualization Comprehension:   A Case Study (Experiment, Study) on Floor Plans
% older: From Pixels to Primitives: Evaluating SVG Decomposition for Enhancing Large Multimodal Models Floor Plan Comprehension

% Decompostion of visualization for Large Multimodal Models undertsanding: Case study for floorplan

% Decomposing "visual elements" for LMM comprehension

% SVG Decomposition for 
% 1. Making visualization machine readable: vector decomposition
% 2. Translating visualization for machine readability 
% 3. From visual to machine readable 

\author{Jeongah Lee}
\affiliation{%
  \institution{University of Massachusetts Amherst}
  \city{Amherst}
  \country{United States}
}
\email{jeongahlee@cs.umass.edu}

\author{Ali Sarvghad}
\email{ali.sarvghad-batn-moghaddam@citystgeorges.ac.uk}
\affiliation{%
  \institution{City St. George's University London}
  \city{London}
  \country{United Kingdom}
}

\renewcommand{\shortauthors}{Lee et al.}

\begin{abstract}
  Large multimodal models (LMMs) are increasingly capable of interpreting visualizations, yet they continue to struggle with spatial reasoning. One proposed strategy is decomposition, which breaks down complex visualizations into structured components. In this work, we examine the efficacy of scalable vector graphics (SVGs) as a decomposition strategy for improving LMMs' performance on floor plans comprehension.  Floor plans serve as a valuable testbed because they combine geometry, topology, and semantics, and their reliable comprehension has real-world applications, such as accessibility for blind and low-vision individuals. We conducted an exploratory study with three LMMs (GPT-4o, Claude 3.7 Sonnet, and Llama 3.2 11B Vision Instruct) across 75 floor plans. Results show that combining SVG with raster input (SVG+PNG) improves performance on spatial understanding tasks but often hinders spatial reasoning, particularly in pathfinding. These findings highlight both the promise and limitations of decomposition as a strategy for advancing spatial visualization comprehension.
  
% Floor plans are essential visualizations that depict spatial organization, functional relationships, and navigational pathways within architectural environments. While large multimodal models (LMMs) have shown strong performance in general visual interpretation tasks, their ability to analyze architectural floor plans remains underexplored. In this study, we evaluate the impact of decomposing floor plan images into vector graphics (SVG) and raster graphics (PNG) on LMMs' performance, using the CubiCasa5K dataset. By testing GPT-4o, Claude 3.7 Sonnet, and Llama 3.2 11B Vision Instruct models, we found that combining PNG and SVG inputs improved spatial understanding tasks—identifying the number of subspaces and their labels—in most cases. However, the impact of decomposition on spatial reasoning tasks—finding a valid and the shortest path—was mixed and often detrimental. These findings highlight the benefits of structured representations for semantic parsing while also revealing persistent limitations in LMMs' spatial reasoning capabilities, especially for tasks requiring a holistic understanding of connectivity.
\end{abstract}



\begin{teaserfigure}
  \includegraphics[width=\textwidth]{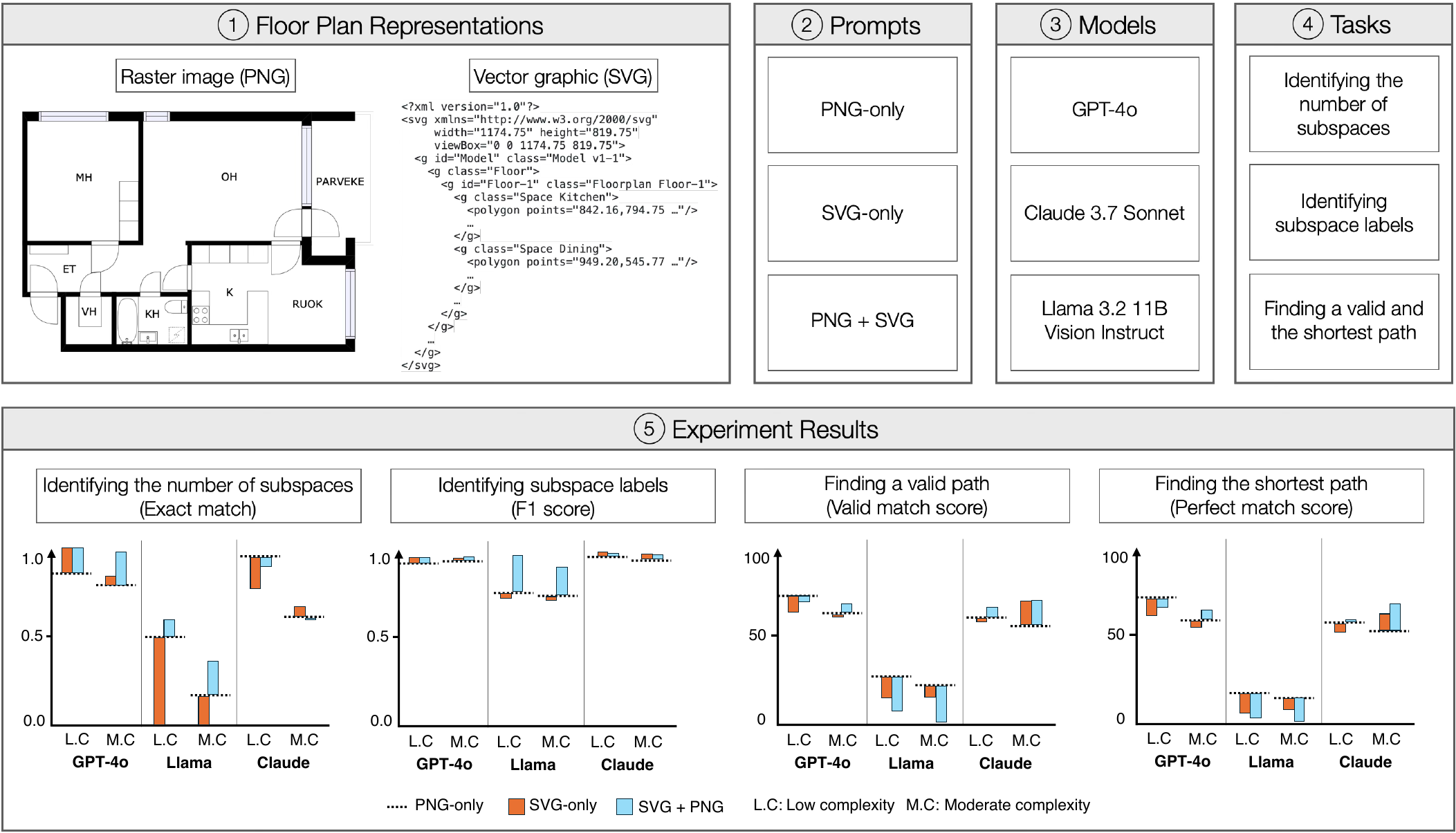}
  \caption{ 
  In this work, we investigated the impact of using vector graphics as an image decomposition strategy on large multimodal models’ comprehension of architectural floor plans. The top row presents a high-level overview of the experimental design (1, 2, 3, and 4). We evaluated 75 floor plans across two complexity levels (33 low and 42 moderate). Using the PNG-only condition as the baseline, we analyzed performance changes under the SVG-only and SVG+PNG conditions. The findings (bottom row) show that combining vector and raster graphics (SVG+PNG) generally improves performance in identifying the number of subspaces and their labels. However, the magnitude of these effects varies depending on the model and the complexity of the floor plan. In contrast, for the pathfinding task, decomposition—whether SVG-only or SVG+PNG—appeared to hinder model performance.
  }
  \Description{}
  \label{fig:teaser}
\end{teaserfigure}

\received{12 September 2025}
\received[revised]{12 December 2025}
\received[accepted]{1 January 2025}

\maketitle
\section{Introduction}
Large multimodal models (LMMs), capable of jointly processing visual and textual information, are becoming increasingly capable of graphics comprehension. Recent studies demonstrate their potential to interpret statistical visualizations~\cite{bendeck2024empirical,li2024visualization,zeng2024advancing}, evaluate encoding quality~\cite{kim2023good}, identify misleading visualizations~\cite{alexander2024can}, and respond to user queries about charts~\cite{masry2022chartqa}. These advances highlight a growing ability of LMMs to support visualization analysis and interaction.

However, despite this progress, LMMs face persistent challenges in reasoning over structured graphics. For example, even state-of-the-art models perform only marginally above chance on multi-step spatial reasoning tasks such as localization and navigation in the TopViewRS benchmark~\cite{li2024topviewrs}, and they often fail on basic relational reasoning (e.g., left/right, above/below)\cite{chenSpatialVLM2024, kamath2023s, liuVisualSpatialReasoning2023}. Similar limitations have been observed in indoor navigation contexts\cite{rajabi2023towards, zhangResearchNavigation2025}. These findings suggest that while LMMs can extract surface-level information, they struggle with deeper geometric and topological understanding.

One promising approach for addressing these challenges is \textit{decomposition}, a method in which complex diagrams are broken down into smaller, semantically coherent components before passing them to a model. Li et al.\cite{lichain}, for instance, demonstrated that representing diagrams as a ``chain-of-regions'' with structured attributes can improve reasoning accuracy. Inspired by this idea, we investigate whether vector decomposition, representing a visualization as scalable vector graphics (SVGs), can improve LMM performance on spatial understanding and reasoning. SVGs encode graphics as XML-based hierarchies of primitives (e.g., \verb|<polygon>| for room boundaries, \verb|<text>| for labels, and \verb|<g>| groups for organization), thereby decomposing a visual scene into discrete, interpretable elements\cite{Lowlevel2024, eisenbergSVGEssentialsProducing2014,caiInvestigation2025,pengRoles2004, Polygraph2025}.

We selected floor plan images as the visualization domain for our experiments. Floor plans' unique integration of geometry, topology, and semantics within a single 2D representation makes them an ideal testbed for evaluating the effectiveness of decomposition ~\cite{FloorplanRestore2025}. Furthermore, assessing models' performance on understanding floor plans holds practical value for the field of AI-centered graphics accessibility.

In an exploratory study, we evaluated the performance of three LMMs (GPT‑4o, Claude 3.7 Sonnet, Llama 3.2 11B Vision Instruct) on spatial understanding and reasoning with floor plans. Using the models' performance on PNG as the baseline, we measured performance changes under SVG and SVG+PNG conditions. This approach follows established practices, as raster formats like PNG have traditionally been used to evaluate LMMs’ performance on visual tasks. The experiment involved 75 floor plan images, categorized into low (33) and moderate (42) complexity levels. We tested each model on two spatial understanding tasks (identifying the number of subspaces and their labels) and two spatial reasoning tasks (finding a valid path and the shortest path from a given subspace to all others). 

We found that the impact of image decomposition on LMMs' performance varies depending on the task and model (Figure~\ref{fig:teaser}). For spatial understanding tasks, the combination of SVG and PNG outperformed the PNG-only baseline in most cases. This effect was particularly pronounced for Llama, which struggled with pure vector input (SVG-only), while GPT-4o showed only modest improvements with SVG-only input. In contrast, spatial reasoning tasks were hindered by SVG-only prompts in most cases, with partial recovery only under SVG+PNG. This recovery was most notable for moderate-complexity floor plans in GPT-4o and Claude. Across all tasks, Llama's performance remained notably lower than other models and showed a consistent drop with PNG+SVG input in the pathfinding task.

\section{Related Work}

\subsection{Computer Vision \& Floor Plan Understanding}
Floor plan understanding has progressed from heuristic-based methods~\cite{Sheraz:2011:IEEE, heras2014statistical, goyal2021knowledge} 
to deep learning approaches~\cite{lvResidentialFloorPlan2021, pizarro2022automatic, Zeng2019Deep}, with CNNs improving room segmentation and wall detection~\cite{pizarro2022automatic}. 
Recent research has developed specialized models for room classification, structural relationship extraction, and architectural element detection~\cite{SAhmed:2012:IEEE, demir2015coupled, DiffPlanner2025}, but these methods typically rely on single-format representations (e.g., raster-only) and task-specific training. Efforts to convert raster floor plans into vector formats or 3D models~\cite{li2024flona, Liu2017RastertoVector} have focused on geometric accuracy rather than semantic interpretation. Our work advances this by prioritizing both structural fidelity and semantic accessibility in the conversion process.

% Comment: Insert a new subsection “2.2 Visualization & LMMs”. For example:
% Prior work shows that LMMs can answer chart-related questions~\cite{masry2022chartqa, bendeck2024empirical}, critique visual encodings~\cite{kim2023good}, and detect misleading designs~\cite{alexander2024can}. Yet, these models often falter on extracting precise values or reasoning about relationships across encodings~\cite{li2024visualization}. Our findings echo this pattern in spatial visualization: LMMs perform well at semantic read-off (counting, labeling) but struggle with relational reasoning (pathfinding). A key contribution of our study is to isolate representation (PNG vs. SVG vs. combined) as a factor in performance, which chart-focused work typically holds constant.

\subsection{Visualization Comprehension by Large Multimodal Models}
Large multimodal models (e.g., GPT‑4V, Gemini) have made impressive strides in interpreting statistical visualizations, including recognizing trends, comparing groups, and critiquing misleading designs~\cite{bendeck2024empirical,li2024visualization,zeng2024advancing}. However, they still struggle with extracting precise values and accurately interpreting complex encodings~\cite{alexander2024can, bendeck2024empirical, zouImplicitAVE2024}. Chart‑specialized models such as UniChart~\cite{masry2023unichart}, ChartPaLI~\cite{carbune2024chart}, and LLaVA‑Chart~\cite{zeng2024advancing} leverage targeted pretraining on chart structure and reasoning traces to push accuracy on benchmarks like ChartQA~\cite{masry2022chartqa} to 80–85\%. 

In contrast, spatial visualizations (e.g., maps and floor plans) remain challenging. Models perform near chance on object localization and multi‑step spatial inference, with only modest (~5–6\%) gains from chain‑of‑thought prompting~\cite{li2024topviewrs}. Early solutions blending computer‑vision region segmentation~\cite{lichain} or massive synthetic spatial data~\cite {li2024topviewrs} illustrate the potential of structured visual representations and domain‑specific pretraining. But fundamental spatial grounding and relational reasoning continue to pose open challenges~\cite{kamath2023s, lichain, majicSpatial2024}.

Recent work by Xing et al.~\cite{xing2025can} introduced MapBench, a benchmark to evaluate LMMs' ability to interpret human-readable pixel-based maps for path-finding. Their contributions focus on defining a formalized navigation graph (MSSG) and a Chain-of-Thought prompting framework tailored to map-space reasoning. 
While their work offers insights into structured reasoning on general maps, our study differs by focusing on architectural floor plans rather than stylized outdoor maps. In addition to route finding, our evaluation includes diverse spatial reasoning tasks such as identifying subspace number and labels. We also systematically examine how input representation formats affect model performance—a dimension largely overlooked in prior work.

Closest to our work, Li et al.~\cite{li2024topviewrs} evaluated ten LMMs on spatial perception tasks, including object and scene recognition and localization, as well as reasoning tasks such as spatial relations, room counting, and navigation. They presented models with photorealistic or semantic top-view floor plans, accompanied by multiple-choice questions, and measured their accuracy. In contrast, our work isolates the effect of input format, raster (PNG), scalable vector graphics (SVG), and their combination, on LMMs' floor plan comprehension. Inspired by the Chain‑of‑Region framework, which suggests that decomposing diagrams into discrete regions and feeding detailed visual metadata into Vision Language Model (VLM) prompts boosts analysis accuracy~\cite{lichain}, we further examine whether SVG's explicit encoding of geometric primitives (precise spatial coordinates, shapes, and sizes) may reduce visual ambiguity and thereby enhance model performance.

\section{Methodology}
\subsection{Test Dataset Preparation}

\begin{figure}[htbp] 
    \centering
    \includegraphics[width=\textwidth]{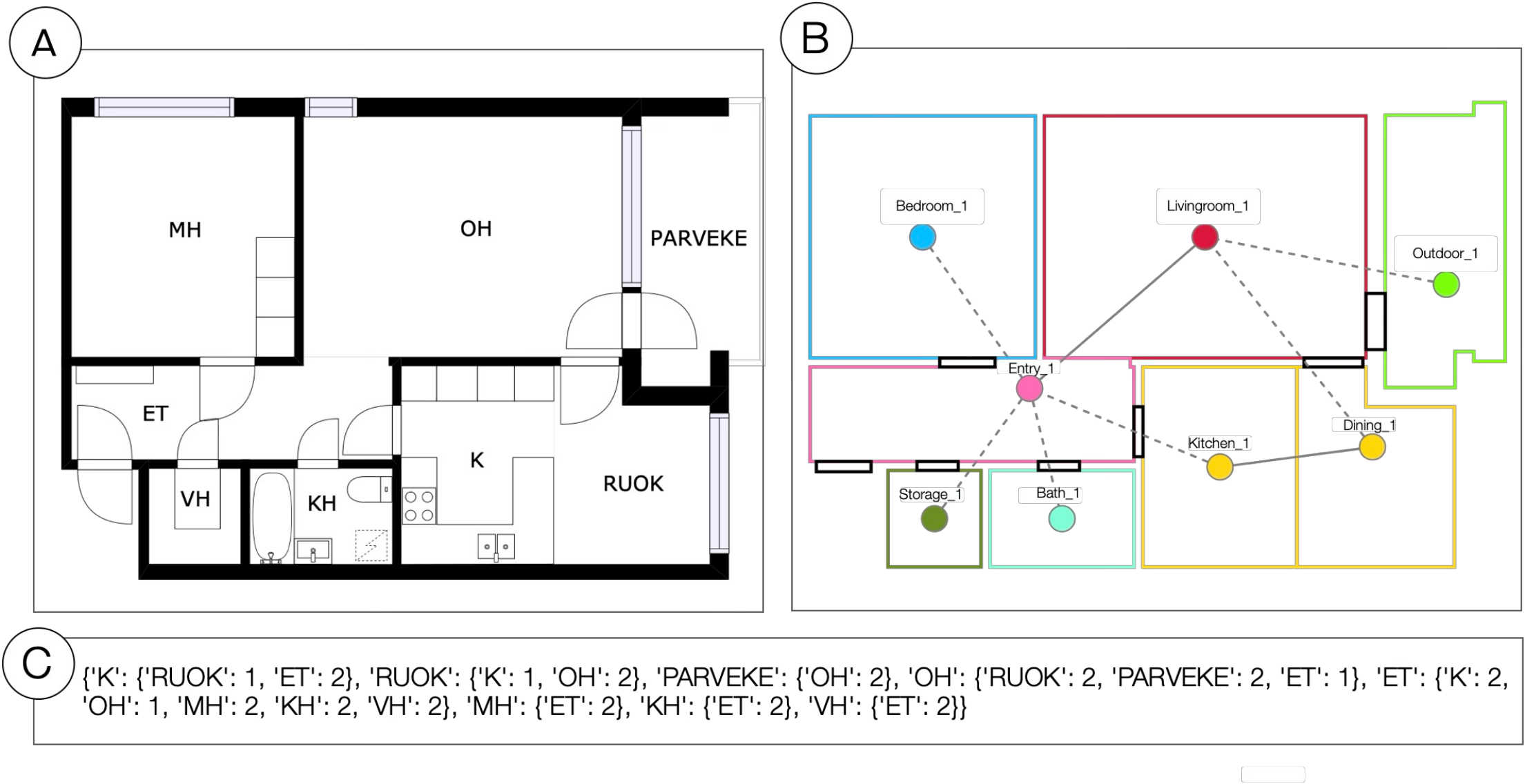} 
    \caption{Path validity evaluation in floor plan graphs. 
    (A) Original floor plan. 
    (B) Graph representation with rooms as nodes; solid lines indicate open-space connections, and dashed lines indicate door-based connections. 
    (C) Adjacency list for path validation; 1 denotes open-space connectivity, and 2 denotes door-based connectivity.
    }
    \label{fig:graph} 
\end{figure}

We curated a test dataset of 75 floor plan images from two complementary sources: \textit{CubiCasa5K}~\cite{kalervo2019cubicasa5k} and \textit{CubiGraph5K}~\cite{lin2021cubigraph5k}. CubiCasa5K comprises 5,000 annotated SVG floor plans, while CubiGraph5K includes corresponding node-link graph representations for each floor plan, modeling subspaces as nodes and their connections as edges—enabling topological reasoning and path-based evaluation (Figure~\ref{fig:graph}). Here, a subspace refers to any enclosed or open functional area within a floor plan, including rooms, utility spaces, circulation areas, and other architecturally defined zones. 

We began with 3,486 single-story floor plans and stratified them by graph-theoretic complexity using three structural indices~\cite{xing2025can}: the Elements Index, measuring the number of subspaces and connections; the Meshedness Index, capturing the density of cycles; and the Average Shortest Path Length Index, reflecting average navigational effort between subspaces. These indices jointly represent spatial scale, redundancy, and traversal difficulty. Based on these metrics, we categorized 1,528 plans as \textit{low} and 1,958 as \textit{moderate}. 

We then applied strict inclusion criteria to retain only 426 high-quality candidates: (1) complete subspace labeling, (2) no “undefined” labels, and (3) no duplicate identifiers. From this set, we selected 75 floor plans (33 \textit{low}, 42 \textit{moderate}) using proportional stratified sampling across quantile bins of normalized complexity scores. Of note, the sample preserved the statistical characteristics of the full dataset, as indicated by a small effect size (Cohen’s $d = 0.28$) between the selected sample and the source population, and maintained the original low-to-moderate ratio (1:1.28)

\subsection{Experiment Design}
% Comment: At the start of 3.2 (tasks) or when reporting results, add:
% Following visualization task taxonomies, our study spans both read-off tasks (entity identification: subspace counting, labeling) and relational reasoning tasks (topology/connectivity: valid path, shortest path).

We conducted the experiment using two commercial models (GPT‑4o and Claude 3.7 Sonnet) and one open-source model (Llama 3.2 11B Vision Instruct). Each model was evaluated on two spatial understanding tasks: (1) identifying the number of subspaces, and (2) identifying subspace labels, as well as two spatial reasoning tasks: (1) finding a valid path from a subspace to all others, and (2) finding the shortest path from a subspace to all others. 

The first two tasks assess models' topological and semantic understanding of the floor plan, while the pathfinding task evaluates their spatial relationship comprehension and connectivity reasoning. For the latter task, each model was queried with all possible start–end subspace pairs, resulting in 384 evaluated paths for the low-complexity set and 598 for the moderate-complexity set. The two pathfinding tasks distinguish models' ability to reason spatially, differentiating between correct reasoning and efficiency in finding the optimal path.

We varied the input type to examine the effects of decomposition on model performance. Specifically, we tested three input types: (1) PNG-only, (2) SVG-only, and (3) PNG+SVG. In all conditions, PNGs were provided as rendered images. SVGs were supplied as raw XML code. While alternative prompting strategies, such as including node-link representations of floor plans or in-context learning, could influence model performance, we did not include them in this study since our goal was to isolate and examine the pure effects of decomposition on the models' performance.

\section{Results}

% ============== Task 1 ==============
\subsection{Identifying the number of subspaces}

\begin{comment}
\begin{table}[htbp]
    \centering
    \caption{Subspace number identification accuracy by floor plan complexity and input type. Each cell shows the mean exact match score.}
    \label{tab:subspace_num}
    \small
    \renewcommand{\arraystretch}{1.0}
    \setlength{\tabcolsep}{7.5pt}
    \begin{tabular}{ll|ccc}  % ← vertical bar after second column
        \toprule
        \multicolumn{2}{c|}{} & \multicolumn{3}{c}{\textbf{Model}} \\
        \cmidrule(lr){3-5}
        \textbf{Complexity} & \textbf{Input Type} & \textbf{GPT-4o} & \textbf{Llama 3.2} & \textbf{Claude 3.7} \\
        \midrule
        \multirow{3}{*}{Low} 
            & png       & 0.85 & 0.50 & 0.95 \\
            & svg       & 1.00 & 0.00 & 0.75 \\
            & png+svg   & 1.00 & 0.60 & 0.90 \\
        \cmidrule(lr){1-5}
        \multirow{3}{*}{Moderate} 
            & png       & 0.75 & 0.17 & 0.61 \\
            & svg       & 0.79 & 0.00 & 0.68 \\
            & png+svg   & 0.96 & 0.39 & 0.60 \\
        \bottomrule
    \end{tabular}
\end{table}
\end{comment}

For this task, we compared each model's prediction for a given floor plan against the ground-truth answer from the CubiGraph5K dataset, using an exact match criterion. A score of one was assigned for each correct match, and zero otherwise. The impact of decomposition varied notably across models (Table~\ref{tab:subspace_num} and Figure~\ref{fig:teaser}). 

GPT-4o showed performance improvements under all tested conditions, with particularly notable gains under the SVG+PNG condition on moderate-complexity floor plans, where accuracy increased by 0.21 points. In contrast, Llama's performance lagged behind the other models under the baseline condition, and decomposition produced mixed results. SVG-only further degraded the model's performance, dropping it to 0.00 on both low and moderate complexity floor plans, while SVG+PNG led to substantial improvements, including a 0.22 point gain on moderate-complexity floor plans. 

Further analysis showed that Llama consistently predicted more rooms than the ground truth, such as an average of 7 additional rooms for low-complexity floor plans, and up to 13 in extreme cases. Claude performed best with PNG-only input on low-complexity floor plans, and in most cases, decomposition led to a slight performance drop; SVG-only input, however, yielded a small performance gain on moderate-complexity floor plans.

\begin{table}[htbp]
    \centering
    \caption{Subspace number identification accuracy by floor plan complexity and input type. Each cell shows the mean exact match score.}
    \label{tab:subspace_num}
    \small
    \renewcommand{\arraystretch}{1.0}
    \setlength{\tabcolsep}{5pt} % 열 간격 조정
    \begin{tabularx}{\textwidth}{ll|XXX} 
        \toprule
        \multicolumn{2}{c|}{} & \multicolumn{3}{c}{\textbf{Model}} \\
        \cmidrule(lr){3-5}
        \textbf{Complexity} & \textbf{Input Type} & \textbf{GPT-4o} & \textbf{Llama 3.2} & \textbf{Claude 3.7} \\
        \midrule
        \multirow{3}{*}{Low} 
            & png       & 0.85 & 0.50 & 0.95 \\
            & svg       & 1.00 & 0.00 & 0.75 \\
            & png+svg   & 1.00 & 0.60 & 0.90 \\
        \cmidrule(lr){1-5}
        \multirow{3}{*}{Moderate} 
            & png       & 0.75 & 0.17 & 0.61 \\
            & svg       & 0.79 & 0.00 & 0.68 \\
            & png+svg   & 0.96 & 0.39 & 0.60 \\
        \bottomrule
    \end{tabularx}
\end{table}

% ================== Task 2=============
\subsection{Identifying subspace labels}

\begin{comment}
\begin{table}[htbp]
    \centering
    \caption{Subspace labels identification accuracy by floor plan complexity and input type. Each cell shows the mean F\textsubscript{1} score.}
    \label{tab:subspace_label}
    \small
    \renewcommand{\arraystretch}{1.0}
    \setlength{\tabcolsep}{7.5pt}
    \begin{tabular}{ll|ccc}  % ← vertical bar after second column
        \toprule
        \multicolumn{2}{c|}{} & \multicolumn{3}{c}{\textbf{Model}} \\
        \cmidrule(lr){3-5}
        \textbf{Complexity} & \textbf{Input Type} & \textbf{GPT-4o} & \textbf{Llama 3.2} & \textbf{Claude 3.7} \\
        \midrule
        \multirow{3}{*}{Low} 
            & png       & 0.96 & 0.71 & 0.97 \\
            & svg       & 1.00 & 0.67 & 0.99 \\
            & png+svg   & 1.00 & 0.96 & 0.98 \\
        \cmidrule(lr){1-5}
        \multirow{3}{*}{Moderate} 
            & png       & 0.97 & 0.70 & 0.95 \\
            & svg       & 0.99 & 0.67 & 0.98 \\
            & png+svg   & 1.00 & 0.88 & 0.97 \\
        \bottomrule
    \end{tabular}
\end{table}
\end{comment}

For this task, we calculated the F\textsubscript{1} score for each floor plan by comparing the model’s predicted subspace labels against the ground-truth labels provided in the CubiGraph5K dataset.

We found that providing both PNG and SVG as input led to performance gains across all models.(Table~\ref{tab:subspace_label} and Figure~\ref{fig:teaser}). While the difference between SVG-only and SVG+PNG was modest for GPT‑4o and Claude, the effect was substantially more pronounced for Llama, with a 0.25 point gain observed on low-complexity floor plans. Floor plan complexity did not notably interact with models' performance across different input types for this task.

\begin{table}[htbp]
    \centering
    \caption{Subspace labels identification accuracy by floor plan complexity and input type. Each cell shows the mean F\textsubscript{1} score.}
    \label{tab:subspace_label}
    \small
    \renewcommand{\arraystretch}{1.0}
    \setlength{\tabcolsep}{5pt} 
    \begin{tabularx}{\textwidth}{ll|XXX} 
        \toprule
        \multicolumn{2}{c|}{} & \multicolumn{3}{c}{\textbf{Model}} \\
        \cmidrule(lr){3-5}
        \textbf{Complexity} & \textbf{Input Type} & \textbf{GPT-4o} & \textbf{Llama 3.2} & \textbf{Claude 3.7} \\
        \midrule
        \multirow{3}{*}{Low} 
            & png       & 0.96 & 0.71 & 0.97 \\
            & svg       & 1.00 & 0.67 & 0.99 \\
            & png+svg   & 1.00 & 0.96 & 0.98 \\
        \cmidrule(lr){1-5}
        \multirow{3}{*}{Moderate} 
            & png       & 0.97 & 0.70 & 0.95 \\
            & svg       & 0.99 & 0.67 & 0.98 \\
            & png+svg   & 1.00 & 0.88 & 0.97 \\
        \bottomrule
    \end{tabularx}
\end{table}

% ================== Task 3=============
\subsection{Finding a valid and the shortest path}

\begin{comment}
\begin{table}[htbp]
    \centering
    \caption{Pathfinding performance by floor plan complexity and input type. Each cell presents the mean VMS and PMS in the format VMS / PMS.}
    \label{tab:pathfinding}
    \small
    \renewcommand{\arraystretch}{1.0}
    \setlength{\tabcolsep}{7.5pt}
    \begin{tabular}{ll|ccc}
        \toprule
        \multicolumn{2}{c|}{} & \multicolumn{3}{c}{\textbf{Model}} \\
        \cmidrule(lr){3-5}
        \textbf{Complexity} & \textbf{Input Type} & \textbf{GPT-4o} & \textbf{Llama 3.2} & \textbf{Claude 3.7} \\
        \midrule
        \multirow{3}{*}{Low} 
            & png       & 74.1 / 72.8 & 25.9 / 19.7 & 61.2 / 59.9 \\
            & svg       & 63.9 / 62.6 & 10.2 / 6.1  & 58.5 / 53.1 \\
            & png+svg   & 69.4 / 67.3 &  5.4 / 3.4  & 65.3 / 61.2 \\
        \cmidrule(lr){1-5}
        \multirow{3}{*}{Moderate} 
            & png       & 62.1 / 62.1 & 23.2 / 16.8 & 54.7 / 53.7 \\
            & svg       & 61.1 / 57.9 & 13.7 / 9.5  & 71.6 / 64.2 \\
            & png+svg   & 68.4 / 67.4 &  1.1 / 1.1  & 72.6 / 69.5 \\
        \bottomrule
    \end{tabular}
\end{table}
\end{comment}

For this task, we evaluated each model's performance by answering two questions for each predicted path: (1) Is the path valid? And, if so, (2) Is it the shortest path?
To answer the first question, we calculated the Valid Match Score (VMS) by checking whether the predicted path adheres to the ground-truth connectivity information provided in the CubiGraph5K dataset. Each floor plan is represented as a graph, where subspaces are nodes and their connections are encoded as edges (Figure~\ref{fig:graph}). 
For the second question, we computed the Perfect Match Score (PMS), which measures the percentage of valid paths that exactly match the ground-truth shortest paths between the same start and end points. PMS serves as a stricter measure of pathfinding accuracy, capturing the proportion of cases where the model identified not just a valid route, but the optimal one.
The effects of decomposition varied across models (Table~\ref{tab:pathfinding} and Figure~\ref{fig:teaser}). GPT-4o's accuracy dropped below the PNG baseline for most conditions, except for PNG+SVG input for moderate-complexity floor plans.

Similar to other tasks, Llama's performance was significantly lower than the other models and was notably degraded with both SVG and SVG+PNG inputs. 
Claude, with the exception of SVG-only on low-complexity floor plans, generally showed performance improvements under other conditions, with particularly high gains for moderate-complexity floor plans. The results for identifying the shortest path closely mirrored the patterns observed for identifying a valid path.

Further analysis of invalid paths revealed two primary error types. The first error was connectivity misinterpretation, where models incorrectly assumed traversability across walls or non-adjacent subspaces, which were common. For example, in Figure~\ref{fig:graph} (A), models incorrectly interpreted that direct movement from RUOK to PARVEKE was possible without passing through intermediate spaces. This error accounted for 100\% of GPT‑4o’s and 92.1\% of Llama’s invalid paths under the SVG-only input, and for 100\% of both models’ invalid paths under the SVG+PNG input. The second error was the omission of essential intermediate rooms, which caused dead-ends, disconnections, or unnecessary backtracking. For example, in Figure~\ref{fig:graph} (A), the valid path from MH to RUOK is “MH, ET, K, RUOK”, but the model incorrectly predicted “MH, ET, RUOK”, skipping the K. This error accounted for 7.9\% of LLaMA 3.2’s invalid paths under the SVG-only input.

\begin{table}[htbp]
    \centering
    \caption{Pathfinding performance by floor plan complexity and input type. Each cell presents the mean VMS and PMS in the format VMS / PMS.}
    \label{tab:pathfinding}
    \small
    \renewcommand{\arraystretch}{1.0}
    \setlength{\tabcolsep}{5pt} % 열 간격 조정
    \begin{tabularx}{\textwidth}{ll|XXX}
        \toprule
        \multicolumn{2}{c|}{} & \multicolumn{3}{c}{\textbf{Model}} \\
        \cmidrule(lr){3-5}
        \textbf{Complexity} & \textbf{Input Type} & \textbf{GPT-4o} & \textbf{Llama 3.2} & \textbf{Claude 3.7} \\
        \midrule
        \multirow{3}{*}{Low} 
            & png       & 74.1 / 72.8 & 25.9 / 19.7 & 61.2 / 59.9 \\
            & svg       & 63.9 / 62.6 & 10.2 / 6.1  & 58.5 / 53.1 \\
            & png+svg   & 69.4 / 67.3 &  5.4 / 3.4  & 65.3 / 61.2 \\
        \cmidrule(lr){1-5}
        \multirow{3}{*}{Moderate} 
            & png       & 62.1 / 62.1 & 23.2 / 16.8 & 54.7 / 53.7 \\
            & svg       & 61.1 / 57.9 & 13.7 / 9.5  & 71.6 / 64.2 \\
            & png+svg   & 68.4 / 67.4 &  1.1 / 1.1  & 72.6 / 69.5 \\
        \bottomrule
    \end{tabularx}
\end{table}

% Comment: And in Results (end of §4):
% Viewed through visualization task taxonomies, SVG+PNG favors read-off tasks (counting, labeling) by exposing discrete primitives, while relational tasks (path validity, optimality) require preserving holistic raster context—explaining the SVG-only drop and PNG+SVG recovery (Fig. 1; Tables 1–3).

% Comment: Discussion → Design implications for AI-ready visualizations. Add a paragraph:
% Our findings suggest strategies for exporting AI-ready visualizations. First, pair structured SVG with a raster overview (“dual-channel export”) to balance semantic precision with global context. Second, encode typeful primitives (e.g., <g> groups for rooms, doorways, or links) to guide model interpretation. Third, include a topology sidecar (graph representation of adjacency) to stabilize path queries, as illustrated in Figure 2. Finally, adopt chunking strategies that combine region-level SVGs with holistic PNGs to mitigate over-fragmentation. These principles extend beyond floor plans to charts (marks and data bindings), maps (layered vectors with base maps), and diagrams (nodes and edges), offering concrete practices for visualization authors seeking to optimize LMM comprehension.

\section{Discussion}
The findings of our experiment suggest that vector decomposition may enhance LMMs' spatial understanding and inference of discrete floor plan regions, likely yielding performance gains for topology-based tasks, such as those tested in our study. However, the benefits of vector decomposition were less consistent and mixed for the spatial reasoning tasks. Unlike identifying the number of subspaces and their labels, finding a valid path or the shortest path requires an additional layer of reasoning—specifically, understanding how subspaces are connected and traversable. It is plausible that while decomposition enhances geometry understanding, it may also cause fragmentation, hindering the model's ability to holistically analyze and infer relational cues from isolated geometric primitives.
Future research could explore combining vector decomposition with other modalities, such as textual descriptions or semantic annotations, to build more comprehensive floor plan representations, enabling more accurate spatial reasoning and decision-making in real-world applications.
%Future research could explore combining vector decompositions with other modalities, such as textual descriptions or semantic annotations, to create more comprehensive representations of floor plans. Multi-modal approaches could further enhance model understanding of complex environments, enabling more accurate spatial reasoning and decision-making in real-world applications.

Performance differences across models—particularly Llama’s difficulty with SVG-only inputs in identifying the number of subspaces—suggest that model architecture plays a critical role in leveraging vector-based decompositions. Models with stronger capabilities in handling symbolic or structured data may benefit more from vector decomposition than those that rely more heavily on pattern recognition. Llama's overestimation of subspace numbers may arise from the model's ineffective understanding of the XML structure, as it hallucinates nonexistent subspaces by arbitrarily interpreting XML tags as subspaces. This issue might be mitigated by developing decomposition-specific grammars that guide the construction of SVGs optimized for model comprehension. 

Our study did not reveal a consistent relationship between floor plan complexity and model performance across tasks. However, the test dataset used in this study does not fully capture the diversity of architectural designs, and future research should explore datasets with a broader range of floor plan complexities. 
Additionally, while manual inspection revealed that the graph-theoretic complexity measure~\cite{xing2025can} yielded reasonable results, the classification relied on fixed threshold values, which were somewhat arbitrary: maps were labeled as \textit{easy} for scores below 0.33, \textit{medium} for scores between 0.33 and 0.66, and \textit{hard} for scores above 0.66.

Building on these findings, while our study provides valuable insights into LMMs’ floor plan comprehension, several important factors were not tested that should be considered in future work. First, our work encompasses a limited set of LMMs, and testing a broader range of models, especially those optimized for different tasks, is necessary to fully assess the potential of decomposition strategies. Spatial understanding and reasoning tasks tested in our study also do not fully capture the complexity of real-world spatial reasoning, and future work should explore more dynamic or multi-step tasks. Lastly, our focus on vector decomposition leaves room for exploring other techniques, such as semantic annotations or dynamic interactions, which could provide a more comprehensive understanding of model performance. Future work in this direction can pave the way for the development of AI systems for accessibility, architectural design, urban planning, or indoor navigation that are efficient in tasks that require spatial reasoning and pathfinding ~\cite{Actfloorgan2023}.

% Comment: Broaden claim (end of Discussion / Conclusion) Add:
% We view decomposition as a visualization pipeline intervention: when authors export graphics, pairing structured primitives with raster overviews can improve semantic comprehension while retaining context for relational reasoning. This approach is not limited to floor plans, but should extend to charts (SVG marks + PNG snapshot), maps (vector layers + base map), and diagrams (node/edge lists + overview).

% Comment: Accessibility & agency (Applications). At the end of Discussion / Implications:
% These insights are particularly relevant for graphics accessibility. Structured SVGs and sidecar graphs enable conversational agents to generate more accurate summaries and navigation cues for blind and low-vision users, while raster overviews support holistic queries and prevent misrouting. By linking decomposition to accessibility, we align with core VIS and HCI concerns about inclusive design.

\section{Limitations}
% Missing baselines and ablations: The paper doesn't compare against other structured representations (e.g., textual descriptions of floor plans, simplified schematic representations) or explore different ways of presenting SVG information (e.g., hierarchical parsing, selective element highlighting).

Our study provides initial insights into the role of vector decomposition in enhancing LMMs’ comprehension of floor plans, but several limitations must be acknowledged.
First, we evaluated only three models (GPT-4o, Claude 3.7 Sonnet, and Llama 3.2 11B Vision Instruct). Other widely used LMMs, such as Gemini and domain-specialized vision–language models, were excluded due to resource and budget constraints. Future work should broaden the scope to include a more diverse set of models with varying architectures and training paradigms .
Second, our focus on floor plans narrows the generalizability of the findings. While floor plans integrate geometry, topology, and semantics and are thus valuable testbeds, the visualization space is much broader. Testing decomposition on other visualization types, such as statistical charts, maps, and scientific diagrams, will be critical for assessing the broader utility of this strategy.
Third, we investigated decomposition only through SVG representations, which decompose visuals into geometric primitives. While this approach highlights structural details, other decomposition strategies, such as graph-based abstractions, semantic annotations, or multimodal encodings, may yield different trade-offs and warrant exploration .
Finally, we focused on static evaluation tasks. Real-world applications, such as supporting blind and low-vision individuals or assisting robotic navigation, often involve interactive, multi-step reasoning under uncertainty. Extending decomposition strategies to such dynamic contexts remains an open challenge.

% Additionally, our dataset included only low- and moderate-complexity floor plans. Architectural designs can vary widely, and extreme complexities, multi-story layouts, or diverse cultural conventions were not represented. Expanding to larger and more diverse datasets will be necessary to fully assess decomposition’s impact.

\section{Conclusion}
Our study examined the impact of decomposing pixel-based images into primitive-based representations on LMMs' floor plan comprehension. We found that combining PNG and SVG improved identifying subspace information, emphasizing the advantages of structured representations for semantic parsing. However, relational reasoning tasks, such as pathfinding, remained challenging and model-dependent, indicating that current LMMs still struggle with holistic spatial understanding. These findings highlight both the potential and limitations of structured visual representations in enhancing model performance. Future research should focus on refining decomposition methods, exploring targeted prompting strategies, and incorporating alternative techniques, such as semantic annotations or multi-modal data, to advance spatial reasoning and structured visual understanding in multimodal models. 

% Comment: Future Work (Evaluation roadmap). Add a short paragraph:
% Future work should establish a cross-visualization benchmark varying representation (PNG, SVG, hybrid) × task type (read-off vs. relational) across charts, maps, and floor plans. Reporting accuracy and error modes (e.g., overcounting rooms vs. overcounting bars) would position representation as a first-class design choice for visualization research.

\bibliographystyle{ACM-Reference-Format}
\bibliography{main}
\end{document}